\documentclass[aps,pra,reprint,superscriptaddress]{revtex4-1}
\usepackage{graphicx}% Include figure files
\usepackage{dcolumn}% Align table columns on decimal point
\usepackage{bm}
\usepackage[latin1]{inputenc}
\usepackage{natbib}
\usepackage{amsmath}
\usepackage{amssymb}
\usepackage{amsfonts}
\usepackage{psfrag}
\usepackage{dsfont}
\usepackage{color}
\usepackage{mathtools}
\begin{document}
%\begin{footnotesize}
% Use the \preprint command to place your local institutional report
% number in the upper righthand corner of the title page in preprint mode.
% Multiple \preprint commands are allowed.
% Use the 'preprintnumbers' class option to override journal defaults
% to display numbers if necessary
%\preprint{}

%Title of paper
\title{Angle-resolved time delay in photoemission}

% repeat the \author .. \affiliation  etc. as needed
% \email, \thanks, \homepage, \altaffiliation all apply to the current
% author. Explanatory text should go in the []'s, actual e-mail
% address or url should go in the {}'s for \email and \homepage.
% Please use the appropriate macro foreach each type of information

% \affiliation command applies to all authors since the last
% \affiliation command. The \affiliation command should follow the
% other information
% \affiliation can be followed by \email, \homepage, \thanks as well.
\author{J. W\"{a}tzel}
\email{jonas.waetzel$@$physik.uni-halle.de}
%\homepage[]{Your web page}
%\thanks{}
\affiliation{Institut f\"{u}r Physik, Martin-Luther-Universit\"{a}t Halle-Wittenberg, K-FH-v-Fritsch-Str.3, 06112 Halle (Saale), Germany}
%\affiliation{}

\author{A. S. Moskalenko}
\affiliation{Institut f\"{u}r Physik,
Martin-Luther-Universit\"{a}t Halle-Wittenberg, K-FH-v-Fritsch-Str.3, 06112 Halle (Saale), Germany}

\affiliation{Institut fur Physik, Universitat Augsburg, D-86135
Augsburg, Germany}

\affiliation{Ioffe Physical-Technical Institute of RAS, 194021
St.~Petersburg, Russia}

\author{Y. Pavlyukh}

 \affiliation{Institut f\"{u}r Physik,
Martin-Luther-Universit\"{a}t Halle-Wittenberg, K-FH-v-Fritsch-Str.3, 06112 Halle (Saale), Germany}

\author{J. Berakdar}
\affiliation{Institut f\"{u}r Physik,
Martin-Luther-Universit\"{a}t Halle-Wittenberg, K-FH-v-Fritsch-Str.3, 06112 Halle (Saale), Germany}

%Collaboration name if desired (requires use of superscriptaddress
%option in \documentclass). \noaffiliation is required (may also be
%used with the \author command).
%\collaboration can be followed by \email, \homepage, \thanks as well.
%\collaboration{}
%\noaffiliation

%\date{\today}

% insert suggested PACS numbers in braces on next line

\date{\today}

\begin{abstract}
 We investigate theoretically the relative time delay of photoelectrons originating from different atomic subshells of noble gases. This quantity was measured via attosecond streaking and studied theoretically by Schultze \emph{et al.} [Science \textbf{328}, 1658 (2010)] for neon. A substantial discrepancy was found between the measured and the calculated values of  the relative time delay. Several theoretical studies were put forward to resolve this issue, e.g., by including correlation effects. In the present paper we explore a further aspect, namely the directional dependence of time delay. In contrast to neon, for argon target a strong angular dependence of time delay is  found near a Cooper minimum.
\end{abstract}

\maketitle

\section{Introduction}

The development of attosecond light sources paved the way for new exciting opportunities for studying time-resolved dynamical processes.
Experiments   on the attosecond electron dynamics in atomic, molecular and condensed matter systems  confirmed the feasibility of this technique for a wide class of systems \cite{Drescher09032001,PhysRevLett.88.173903,YakolevEtAl,QuereEtAl,Sansone20102006,maquet,Yakovlev2}.
Thereby, an  attosecond streaking metrology is employed for tracing the dynamics \cite{Drescher2,Goulielmakis,Cavalieri,Schultze}:
An extreme ultraviolet (XUV) pulse with a duration of a few hundred attoseconds plays the role of the pump and a phase-controlled few-cycle infrared (IR) pulse acts as the probe pulse.
 The coherent XUV pulse is characterized by a moderate intensity, short wavelength and a Keldysh parameter $\gamma\gg1$. Valence electrons, which are emitted due to the action of the XUV field, are accelerated to different final momenta in the presence of the IR field. The  asymptotic momenta depend on the vector potential of the IR-field  while
  the electron recedes from the residual ion allowing thus to access time information via momentum measurements.
  % In a recent study \cite{AndreyPRA} we showed theoretically that on a sub-femtosecond  time scale photonic properties (of fullerene in our example) are quantitatively different and can be revealed with the attosecond metrology.\\
The  experiment in Ref.\cite{Schultze} reports on the time delay in the photoemission  of electrons from the $2p$ subshell relative to those from the $2s$ subshell.
A series of theoretical works were developed to quantitatively reproduce the measurements (for a review of the time delay concept we refer to Ref. \onlinecite{maquet} and references therein). In the present work we refer
For details of the time delay theory/interpretation we refer to the review articles \cite{deCarvalho200283,C3FD00004D}.
The experiment yielded  a relative time delay of $21\pm5$~as. Within an independent electron model Schultze \emph{et al.} were able to calculate a delay of 6.37~as from matrix elements obtained by means of the state-specific expansion approach \cite{Schultze}. Ivanov and Kheifets reached a delay of 6.2 as by calculating scattering phases and dipole matrix elements in the Hartree-Fock (HF) approximation \cite{Ivanov1}. Accounting for first order correlation correction via the random phase approximation with exchange (RPAE)  added 2.2 as. Another factor that may influence the time delay is the interaction between the photoelectron and the infrared  field. The measured delays can be generally divided in two parts. The first is the Wigner-like delay $\tau_{\rm W}$ originating from the XUV photoionization process \cite{Wigner1}, while the second term is a contribution from the interaction of the moving photoelectron (in the Coulomb-field of the residual ion) with the IR field, which is called Coulomb-laser coupling delay $\tau_{\rm CLC}$ \cite{Nagale1,Dahlstrom,Klunder2001}:
\begin{equation}
\tau=\tau_{\rm W}+\tau_{\rm CLC}.
\label{gendelay}
\end{equation}
Zhang and Thumm showed for then one-dimensional hydrogen model that the resulting delay is independent of the IR pulse intensity \cite{ZhangThumm2}. Nagele \emph{et al.} confirmed this statement with the help of three-dimensional full numerical propagation simulations. They found that the Wigner-delay is accessible by streaking only if distortions of both the initial-state entrance channel and the final-state exit channels (e.g. initial state polarization by IR or combined Coulomb-IR fields) are precisely accounted for \cite{Nagale1}. Recently Moore \emph{et al.} \cite{Moore} found a value of $10.2\pm1.3$~as for relative delay between the $2p$ and $2s$ subshells when using the R-Matrix method. The discrepancy between theory and experiment is not yet resolved, however.\\
Recently the time delay studies moved from neon to argon, where interesting phenomena at energies around the Cooper minimum \cite{Guenot2012,Klunder2001,Kheifets2013} are found. Kl\"under \emph{et al}. reported on the relative time delay between the photoemissions from the $3s$ and $3p$ subshells of argon for photon energies in the range of 34 to 40 eV \cite{Klunder2001}. This measurement was repeated later by Gu\'enot \emph{et al.}\cite{Guenot2012}, whose results agree very well, except for that obtained at the highest energy corresponding to 40~eV.\\
The angular dependence of the time delay has received relatively less attention.
E.g., it was calculated for hydrogen atom showing a strong dependence of the time delay around $\pm 90^{\circ}$ with respect to the laser polarization axis \cite{Ivanov2}.
H$_2$ was studied in Ref.\cite{serov}.\\
In the experiment the time-of-flight detector (TOF) collects all electrons within a certain solid angle. We introduce the angle $\theta_{{\boldsymbol{k}}}$ between the asymptotic direction of the momentum of the photoelectron and the laser polarization axis. The latter is chosen to be parallel to the $z$-axis. The aim here is to investigate to which extent the angular dependence influences the resulting time delay with reference to its value by $\theta_{{\boldsymbol{k}}}=0^{\circ}$.
We analyze whether the calculations allow for some trends in the angle-dependent time delay.
To address this point we study  all possible propagation directions of the photoelectrons and calculate the corresponding angle-resolved time delays for neon and argon. \\
We concentrate on the angular dependence of the Wigner time delay $\tau_{\rm W}$, which enters the time delay in eq. \eqref{gendelay} and the delay measured in attosecond streaking experiments. The corrections to the Wigner time delay through the Coulomb-laser coupling effect can be evaluated according to \cite{Dahlstrom2,Nagele2}.
Neon was chosen  because of the experiment by Schultze \emph{et al.} \cite{Schultze} and the large number of theoretical results for the relative time delay between photoemissions from the  $2p$ and $2s$ states;  argon on the other hand is interesting due to the presence of a Cooper minimum in the spectra.
%because of the comparable electronic structure and calculate the angular dependent relative time delay between photoemissions from $3p$ and $3s$ states. To include correlation we %employ the RPAE method as a more complex HF theory \cite{Amusia1,Amusia2}.
\section{Theoretical Model}
The atomic units will be used throughout the text unless indicated otherwise.
The wave function representing a photoelectron wave packet is given by
\begin{equation}
\Phi(\boldsymbol{r},t)=\int{\rm
d}\boldsymbol{k}\,a(\boldsymbol{k},t)\varphi^{(-)}_{\boldsymbol{k}}(\boldsymbol{r})e^{-i\varepsilon_{k}t},
\end{equation}
where $\varphi^{(-)}_{\boldsymbol{k}}(\boldsymbol{r})$ stand for a set of continuum wave functions of the system and $a_{{}}(\boldsymbol{k},t)$ are the corresponding projection coefficients. The projection coefficients corresponding to photoionization of a bound state indexed ${\rm i}$ are in general angle-dependent and can be evaluated as
\begin{equation}
a_{\rm i}(\boldsymbol{k})=-i\int_{-\infty}^{\infty}{\rm d}t'\,\langle\varphi_{\boldsymbol{k}}^{(-)}|{H}_{\rm
int}|\Psi_{\rm i}\rangle e^{i(\varepsilon_{k}-\varepsilon_{\rm i})t'}.
\label{PT}
\end{equation}
 We employ the length gauge $H_{\rm int}=zE(t)$ for the interaction with the laser electric field $E(t)$. Thus, the
 matrix element $D_{\rm i}(\boldsymbol{k})=\langle\varphi_{\boldsymbol{k}}^{(-)}|\hat{z}|\Psi_{\rm i}\rangle$ for transition from the bound state $|\Psi_{\rm i}\rangle$ to the continuum state $|\varphi_{\boldsymbol{k}}^{(-)}\rangle$ with the free energy $\varepsilon_k=k^2/2=\omega_{\rm XUV}+\varepsilon_{\rm i}$ have to be calculated, where $\omega_{\rm XUV}$ is the frequency of XUV pulse and $\varepsilon_{\rm i}$ is the energy eigenvalue of the bound state.
 We express the wave function of the initial state as $\Psi_{\rm i}(\boldsymbol{r})=R_{n_{\rm i},\ell_{\rm i}}(r)Y_{\ell_{\rm i},m_{\rm i}}(\Omega_r)$, where $R_{n_{\rm i}\ell_{\rm i}}(r)$ and $Y_{\ell_{\rm i},m_{\rm i}}(\Omega_{\boldsymbol{r}})$ are the radial and spherical parts of the wave functions indexed by the usual quantum numbers $\ell_{\rm i}$ and $m_{\rm i}$. For the continuum states
 we expand in  partial waves \cite{Dahlstrom,Amusia1,Amusia2}
\begin{equation}
\begin{split}
\varphi^{(-)}_{\boldsymbol{k}}(\boldsymbol{r})=\sqrt{(2\pi)^{3}}\sum_{\ell=0}^{\infty}\sum_{m=-\ell}^{+l}&i^{\ell}R_{k\ell}(r) e^{-i\delta_\ell(k)}\\
&\times Y_{\ell m}^{*}(\Omega_{\boldsymbol{k}})Y_{\ell m}(\Omega_{\boldsymbol{r}}).
\label{PWE}
\end{split}
\end{equation}
The radial wavefunctions are normalized  as $\langle R_{k\ell}|R_{k'\ell}\rangle=\delta(\varepsilon_k-\varepsilon_{k'})$.
For  bound states we use a self-consistent HF method \cite{Chernysheva197657}, while the continuum states are evaluated in the HF frozen core approximation \cite{Chernysheva197987}.
Furthermore, we introduce the scattering phases which are given by $\delta_{\ell}(k)=\sigma_{\ell}(k)+\eta_{\ell}(k)$, where $\sigma_{\ell}(k)={\rm arg}\left[\Gamma(\ell+1-i/k)\right]$ is the Coulomb phase shift \cite{joachain}.  $\eta{_\ell}(k)$ is a phase correction  originating from the short range deviation of the atomic potential from  pure Coulomb potential \cite{Dahlstrom}.\\
Using the partial wave expansion and performing the angular integration we obtain
for the matrix element $D_{\rm i}^{\rm HF}(\boldsymbol{k})$ in the HF approximation:
\begin{equation}
\begin{split}
D_{\rm i}^{\rm HF}(\boldsymbol{k})&=\sqrt{\frac{(2\pi)^{3}}{k}}\sum_{\substack{\ell=\ell_{\rm i}\pm1}}i^{-\ell}e^{i\delta^{\rm HF}_{\ell}(k)}Y_{\ell m_{\rm i}}(\Omega_{\boldsymbol{k}})\\
&~~~\times\begin{pmatrix}\ell&1&\ell_{\rm i}\\m_{\rm i}&0&m_{\rm i}\end{pmatrix} d_{\ell,n_{\rm i}\ell_{\rm i}}(k).
\label{eq:DipoleMatrixElement}
\end{split}
\end{equation}
The reduced dipole matrix element \cite{Amusia1,Amusia2} are given by
\begin{equation}
\begin{split}
d_{\ell,n_{\rm i}\ell_{\rm i}}(k)=&\sqrt{(2\ell+1)(2\ell_{\rm i}+1)}\begin{pmatrix}\ell&1&\ell_{\rm i}\\0&0&0\end{pmatrix}\\
&\times\int_0^\infty{\rm d}r\,r^3R_{k\ell}(r)R_{n_{\rm i},\ell_{\rm i}}(r).
\end{split}
\label{ReducedMatrixElement}
\end{equation}
To include electronic correlation effects to first order one may use RPAE.  Then, the reduced dipole matrix element $d_{\ell,n_{\rm i}\ell_{\rm i}}(k)$ is replaced by a screened matrix element $D_{\ell,n_{\rm i}\ell_{\rm i}}(k)$ which accounts for correlation between the various valence subshells. This matrix element is defined by the self-consistent equation \cite{Amusia1,Amusia2}
\begin{equation}
\begin{split}
D_{\ell,n_{\rm i}\ell_{\rm i}}(k)=&d_{\ell,n_{\rm i}\ell_{\rm i}}(k)+\lim\limits_{\epsilon\rightarrow 0^{^+}}\sum_{n_{\rm j}\ell_{\rm j}}^{\rm occ}\,\,\,\,\,
\mathclap{\displaystyle\int\limits_{l'}} \mathclap{\textstyle\sum}\;\;\;{\rm d}k'\,k'\\
&\times\left[\frac{D_{\ell',n_{\rm j}\ell_{\rm j}}(k')\,\langle n_{\rm j}\ell_{\rm j},k\ell||V||k'\ell',n_{\rm i}\ell_{\rm i}\rangle}
{\omega_{\rm XUV}-\varepsilon_{k'}+\varepsilon_{\rm j}+i\epsilon}\right.\\
&\left.+~\frac{D_{n_{\rm j}\ell_{\rm j},\ell'}(k')\,\langle k'\ell',k\ell||V||n_{\rm j}\ell_{\rm j},n_{\rm i}\ell_{\rm i}\rangle}
{\omega_{\rm XUV}+\varepsilon_{k'}-\varepsilon_{\rm j}}\right].
\end{split}
\end{equation}
The indices ${\rm i}$ and ${\rm j}$ stand for  valence orbitals and  sum/integral sign indicates summation over discrete excited states with
energies $\varepsilon_{k'}=\varepsilon_{n'\ell'}$ as well as integration over the continuum states with the energy $\varepsilon_{k'}=k'^2/2$. The reduced Coulomb matrix elements $\langle n_{\rm j}\ell_{\rm j},k\ell||V||k'\ell',n_{\rm i}\ell_{\rm i}\rangle$ and $\langle k'\ell',k\ell||V||n_{\rm j}\ell_{\rm j},n_{\rm i}\ell_{\rm i}\rangle$ describe the time-forward and time-reversed correlation processes, which include both direct and exchange parts \cite{Amusia1,Amusia2}.
Important for the phase information is that the integration in the time-forward term contains a pole with the consequence that the reduced matrix element is complex and acquires therefore an extra phase. Thus, $D_{\ell,n_{\rm i}\ell_{\rm i}}(k)$ can be expressed as $|D_{\ell,n_{\rm i}\ell_{\rm i}}(k)|e^{i\delta_{\ell}^{\rm RPAE}}$. The RPAE dipole matrix element has now the form
\begin{equation}
\begin{split}
D_{\rm i}^{\rm RPAE}(\boldsymbol{k})=&\sqrt{\frac{(2\pi)^{3}}{k}}\sum_{\substack{\ell=\ell_{\rm i}\pm1}}i^{-\ell}e^{i\left(\delta^{\rm HF}_{\ell}(k)+\delta^{\rm RPAE}_{\ell}(k)\right)}\\
&\times\begin{pmatrix}\ell&1&\ell_{\rm i}\\m_{\rm i}&0&m_{\rm i}\end{pmatrix}Y_{\ell m_{\rm i}}(\Omega_{\boldsymbol{k}})|D_{\ell,n_{\rm i}\ell_{\rm i}}(k)|.
\label{eq:DipoleMatrixElementRPA}
\end{split}
\end{equation}
The Wigner time delay is defined as the energy derivative of photoionization amplitude of the respective subshell i, i.e.
\begin{equation}
\tau^{\ell_{\rm i}m_{\rm i}}_{\rm W}(\theta_{\boldsymbol{k}})=
%\frac{\partial}{\partial E}{\rm arg}\left[f_{\rmi}(\boldsymbol{k})\right]\equiv
\frac{\partial}{\partial E}\mu(\boldsymbol{k}),
\label{GeneralTD}
\end{equation}
where $\mu(\boldsymbol{k})={\rm arg}\left[D_{\rm i}^{\rm RPAE}(\boldsymbol{k})\right]$, which provides a direct connection between the time delay and scattering phases $\delta^{\rm HF}_\ell(k)+\delta^{\rm RPAE}_\ell(k)$, which strongly influence the dipole matrix element \eqref{eq:DipoleMatrixElementRPA}. By writing the spherical harmonics $Y_{\ell m}(\Omega_{\boldsymbol{k}})$ as $N_{\ell}^m P^{m}_\ell(\cos\theta_{\boldsymbol{k}})e^{im\varphi_{\boldsymbol{k}}}$  the phase of the dipole matrix element is cast as
\begin{equation}
\mu(\boldsymbol{k})={\rm atan}\left[\frac{\sum_{\ell=\ell_{\rm i}\pm1}S_{\ell}(\boldsymbol{k})\sin(\phi_{\ell}(\boldsymbol{k}))}
{\sum_{\ell=\ell_{\rm i}\pm1}S_{\ell}(\boldsymbol{k})\cos(\phi_{\ell}(\boldsymbol{k}))}\right],
\label{eq:DME_phase}
\end{equation}
where we used
\begin{equation}
\begin{split}
S_{\ell_{\rm i}\pm1}(\boldsymbol{k})=&\begin{pmatrix}\ell_{\rm i}\pm1&1&\ell_{\rm i}\\m_{\rm i}&0&m_{\rm i}\end{pmatrix}D_{\ell_{\rm i}\pm1,n_{\rm i}\ell_{\rm i}}(k)\\
&\times N_{\ell_{\rm i}\pm1}^{m_{\rm i}}P_{\ell_{\rm i}\pm1}^{m_{\rm i}}(\cos(\theta_{\boldsymbol{k}}))
\end{split}
\end{equation}
and
\begin{equation}
\phi_{\ell_{\rm i}\pm1}(\boldsymbol{k})=\delta^{\rm HF}_{\ell_{\rm i}\pm1}(k)+\delta^{\rm RPAE}_{\ell_{\rm i}\pm1}(k)-(\ell_{\rm i}\pm1)\pi/2+m_{\rm i}\varphi_{\boldsymbol{k}}.
\end{equation}
Equation \eqref{eq:DME_phase} in this form is only suitable for the photoemission from a $np$ state with $\ell_{\rm i}=1$ and $m_{\rm i}=0$. One can see immediately that in this case the phase of the dipole matrix element $\mu(\boldsymbol{k})$ is only dependent on the angle $\theta_{\boldsymbol{k}}$ but not on $\phi_{\boldsymbol{k}}$ (since $m_{\rm i}$ is zero). This means that the time delay $\tau_{\rm W}$ as the energy derivative has an angular dependence which is strongly influenced by the ratio $S_{\ell_{\rm i}+1}/S_{\ell_{\rm i}-1}$ and the scattering phases $\delta^{\rm HF}_{\ell_{\rm i}\pm1}(k)+\delta^{\rm RPAE}_{\ell_{\rm i}\pm1}(k)$. \\
In the cases of photoemission from a $ns$ state ($\ell_{\rm i}=0$ and $m_{\rm i}=0$) and photoemission from a $np$ state with $m_{\rm i}=\pm1$ eq. \eqref{eq:DME_phase} reduces to $\mu(\boldsymbol{k})=\phi_{\ell_{\rm i}+1}(\boldsymbol{k})$, because $S_{\ell_{\rm i}-1}(\boldsymbol{k})$ vanishes. In these special cases the time delay as the energy derivative of $\mu(\boldsymbol{k})$ is only characterized by the properties of the scattering phase $\delta^{\rm HF,RPAE}_{\ell=1}(k)$ or $\delta^{\rm HF,RPAE}_{\ell=2}(k)$, respectively. Therefore, the HF and RPAE theory predicts that we find no angular dependence of these state-specific time delays.\\
The time delays defined as the energy derivative of the phase of the photoionization amplitude eq. \eqref{GeneralTD} are in general energy-dependent. To find the characteristic time delay corresponding to the photon frequency $\hbar\omega_{\rm XUV}$ we have to average over the ionization probability $w_{\rm i}(\varepsilon_k,\theta_{\boldsymbol{k}})=\left|a_{\rm i}(\boldsymbol{k})\right|^2$, which has also an angular dependence due to the angular dependence of $a_{\rm i}(\boldsymbol{k})$. Finally we define the time delays corresponding to photoionization from the $ns$ and $np$ subshells:
\begin{equation}
\begin{split}
&\tau_{\rm W}^{ns}(\theta_{{\boldsymbol{k}}})=\frac{\int{\rm
d}\varepsilon\,w_{\ell_{\rm i}=0,m_{\rm i}=0}(\varepsilon,\theta_{{\boldsymbol{k}}})\tau^{\ell_{\rm i}=0,m_{\rm i}=0}_{{\rm
W}}(\varepsilon,\theta_{{\boldsymbol{k}}})}{\int{\rm d}\varepsilon\,w_{\ell_{\rm i}=0,m_{\rm i}=0}(\varepsilon,\theta_{{\boldsymbol{k}}})}\\
&\tau_{\rm W}^{np}(\theta_{{\boldsymbol{k}}})=\frac{\int{\rm
d}\varepsilon\,\sum_{m_{\rm i}=-1}^{1}w_{\ell_{\rm i}=1,m_{\rm i}}(\varepsilon,\theta_{{\boldsymbol{k}}})\tau^{\ell_{\rm i}=1,m_{\rm i}}_{{\rm
W}}(\varepsilon,\theta_{{\boldsymbol{k}}})}{\int{\rm d}\varepsilon\,\sum_{m_{\rm i}=-1}^{1}w_{\ell_{\rm i}=1,m_{\rm i}}(\varepsilon,\theta_{{\boldsymbol{k}}})}.
\label{AveragedWD}
\end{split}
\end{equation}

\section{Angular dependence of the time delay of neon}

The considered XUV field is modeled as
\begin{equation}
E_{{\rm XUV}}(t)=E_{{0}}\cos^2\left(\frac{\pi t}{2T_{{\rm XUV}}}\right)\cos(\omega_{{\rm XUV}}t)
\end{equation}
for times $t$ within the interval $[-T_{{\rm XUV}},T_{{\rm XUV}}]$
and vanishes otherwise. In view of the experiment \cite{Schultze} the parameter $T{_{\rm XUV}}$ for Ne is chosen such that  the pulse FWHM is 182 as. The amplitude of the electric field of  XUV field is 0.12~a.u. which corresponds to a peak intensity of $5\times10^{14}$ W/cm$^2$.\\
\begin{figure}[t!]
\centering
\includegraphics[width=8cm]{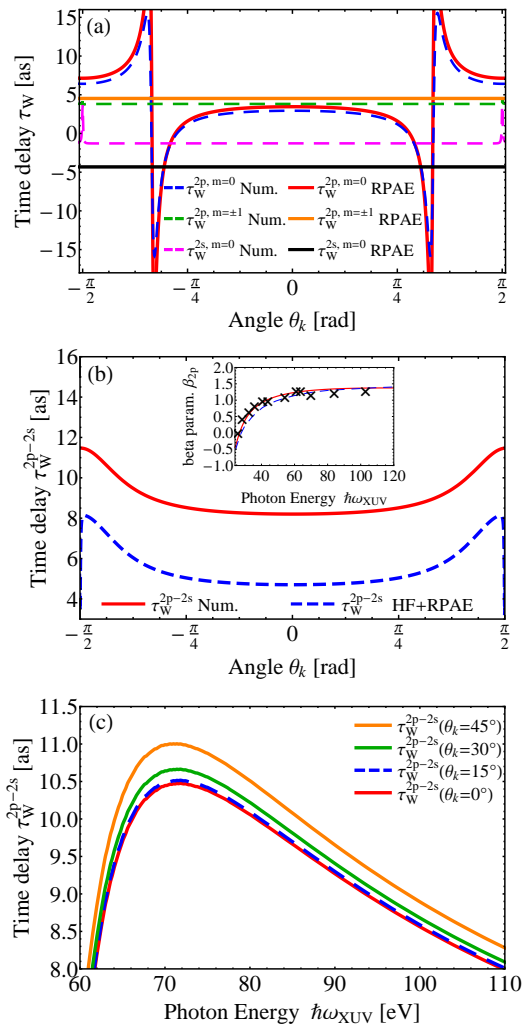}
\caption{(Color online) (a)  energy-dependence of the time delays $\tau_{\rm W}$ for different emission angles $\theta_{\boldsymbol{k}}$ of the photoelectron.
 Results of RPAE calculations and full numerical simulation are shown. (b) angular dependence of the relative delay $\tau_{\rm W}^{2p-2s}$ at $\hbar\omega_{\rm XUV}=106$~eV with contributions of all possible initial states. (c) RPAE relative time delay in dependence on the photon energy for different asymptotic directions $\theta_{\boldsymbol{k}}$.}
\label{DelayNeon}
\end{figure}
In panel (a) of fig.~(\ref{DelayNeon}) we show explicitly the time delays corresponding to the $2s$ and the three possible $2p$ initial states of neon in dependence on the angle $\theta_{\boldsymbol{k}}$ for a photon energy of 106~eV, which is the used frequency in the experiment of Schultze \emph{et al.} \cite{Schultze}. Although the RPAE is a more advanced theory because it treats intershell correlation we compare its predictions of the angular dependence with a full three dimensional SAE (single active electron) propagation. Previous studies showed that in Ne the intershell correlation has no significant impact on the $2p$ photoionization process \cite{Ivanov1,Kheifets2013}, which is according to eq.~\eqref{eq:DME_phase} the origin of the angular dependence of the relative time delay $\tau^{2p-2s}_{\rm W}$. Thus we can compare both results qualitatively. The numerical simulation is realized with the matrix iteration method \cite{Nurhuda}. As an atomic potential for Ne we use  the optimized effective single-particle potential as in Ref.~\cite{Sarsa2004163}. The time dependent wave function is expanded in spherical harmonics, i.e. $\Psi(\boldsymbol{r},t)=\sum_{\ell=0}^{L_{\rm max}}\sum_{m=-\ell}^{\ell}R_{\ell}(r) Y_{lm}(\Omega_{\boldsymbol{r}})$. For $t\rightarrow-\infty$ we define that $\Psi(\boldsymbol{r},t)=\Psi_{\rm i}(\boldsymbol{r})$. Thus, every initial state is propagated from $t=-T_{{\rm XUV}}$ to $T_{{\rm XUV}}$ in the presence of the laser field. After the photelectron wave packet is fully formed, the solution $\Psi(\boldsymbol{r},t>T_{\rm XUV})$ is projected on a set of field-free continuum wave functions $\varphi^{-}_{\boldsymbol{k}}(r)$ and we obtain the projection coefficients $a_{\rm i}(\boldsymbol{k})$ [compare with eq.~\eqref{PT}] corresponding to the bound state $\rm i$, which can be further analyzed to extract the time delay information.\\
The RPAE results indicate a strong angular dependence of the time delay which corresponds to the $2p$ photoionization channel with $m_{\rm i}=0$. However the time delays corresponding to the $2s$ and $2p$ with $m_{\rm i}=\pm1$ channels show no angular dependence. Similar characteristics applies to the results of the full numerical propagation, where the delays corresponding to $\{\ell_{\rm i}=0, m_{\rm i}=0\}$ ($2s$) and $\{\ell_{\rm i}=1, m_{\rm i}=\pm 1\}$ ($2p$) are nearly independent on the angle $\theta_{\boldsymbol{k}}$ in the area between $\pm90^{\circ}$.
We find only a weak angular dependence around $\pm90^{\circ}$, which can be explained by the fact that the partial waves with $\{\ell=1,m=0\}$ and $\{\ell=2,m=\pm 1\}$, which dominate the ionization channels of these initial states, vanish and partial waves with higher angular momenta become decisive.\\
The time delay of the photoionization process corresponding to the $2p$ initial state with $\{\ell_{\rm i}=1, m_{\rm i}=0\}$ shows substantial variations.
The pronounced sharp structures around $\pm57^\circ$ emerge due to the vanishing contribution of the typically dominating transition to the $\ell=2$ partial wave \cite{ArtFano}, that means the corresponding phase $\delta_{\ell=2}(k)=\delta^{\rm HF}_{\ell=2}(k)+\delta^{\rm RPAE}_{\ell=2}(k)$ mainly determines the time delay. From eq.~\eqref{eq:DME_phase} follows that the phase $\mu(\boldsymbol{k})$ is then mainly determindes by the scattering phase $\delta_{\ell=0}(k)=\delta^{\rm HF}_{\ell=0}(k)+\delta^{\rm RPAE}_{\ell=0}(k)$, whose energy derivative is negative. The RPAE calculations and full numerical simulation deliver comparable predictions of the angular dependence of the $2p$ time delay corresponding to $\{\ell_{\rm i}=2, m_{\rm i}=0\}$.\\
While the effect of the RPAE on the delays related to photoionization from the $2p$ subshell is subsidiary \cite{Kheifets2013}, we find that the energy derivative of the addtional phase $\delta^{\rm RPAE}_{\ell=1}(k)$  has a significant influence on the resulting delay $\tau_{\rm W}^{2s}$. This leads to the observed discrepancy between the RPAE results and the full numerical simulation regarding the  relative time delay  including the contributions of all four possible initial states. The panel~(b) of fig.~(\ref{DelayNeon}) shows the angular dependence of the relative delay. The results illustrate that the strong effect of the angular dependence  corresponding to the $2p$ initial state with $\{\ell_{\rm i}=1, m_{\rm i}=0\}$ is nearly compensated by the constant time delays corresponding to $\{\ell_{\rm i}=1, m_{\rm i}=\pm1\}$; underpinning the fact that the sum in the denominator of eq. \eqref{AveragedWD} can be expressed by $\sum_{m_{\rm i}=-1}^{1}w^{l_{\rm i}=1,m_{\rm i}}(\varepsilon_k,\theta_{{\boldsymbol{k}}})\propto1+\beta_{2p}(\varepsilon_k)P_{2}(\cos\theta_{{\boldsymbol{k}}})$, which describes the angular dependence of the photoionization process. From the inset in the panel~(b) we know that $\beta_{2p}(\varepsilon_k)$ is approximately 1.5 around $\hbar\omega_{\rm XUV}=106$~eV, meaning that $1/\left(1+\beta_{2p}(\varepsilon_k)P_{2}(\cos\theta_{{\boldsymbol{k}}})\right)$ has two maxima at $\pm90^{\circ}$.  This explains why the relative time delay grows slowly when approaching $\theta_{\boldsymbol{k}}=\pm90^\circ$. Thus the numerator of eq.~\eqref{AveragedWD} which is determined by $\sum_{m_{\rm i}=-1}^{1}w^{l_{\rm i},m_{\rm i}}(\varepsilon_k,\theta_{\boldsymbol{k}})\tau_{\rm W}^{\ell_{\rm i}m_{\rm i}}$ shows a very weak angular dependence. The value of the relative time delay $\tau^{2p-2s}_{\rm W}$ in the forward direction $\theta_{\boldsymbol{k}}=0^\circ$ is 4.82~as, which is comparable to the the 4.5~as from SAE simulations performed by Schultze \emph{et al.} \cite{Schultze}. The value for the RPAE calculations is 8.19~as, which is in good agreement with the results of Kheifets \emph{et al.} \cite{Ivanov1,Kheifets2013} and serves as a good check for our calculations.\\
The panel~(c) of fig.~(\ref{DelayNeon}) shows the relative time delay of neon as a function of the XUV photon energy for different asymptotic directions $\theta_{\boldsymbol{k}}$ up to  $45^\circ$, exhibiting a very smooth angular dependence. An averaging of the relative time delay over the interval $\theta_{\boldsymbol{k}}\in[-45^\circ,45^\circ]$ at a photon energy of 106~eV leads to a  relative time delay of $\tau^{2p-2s}_{\rm W}=8.41$~as while a smaller acceptance angle of $\theta_{\rm max}=20^\circ$ results in a time delay of $\tau^{2p-2s}_{\rm W}=8.24$~as.
\section{Angular dependence of the time delay of argon}
\begin{figure}[t!]
\centering
\includegraphics[width=8.75cm]{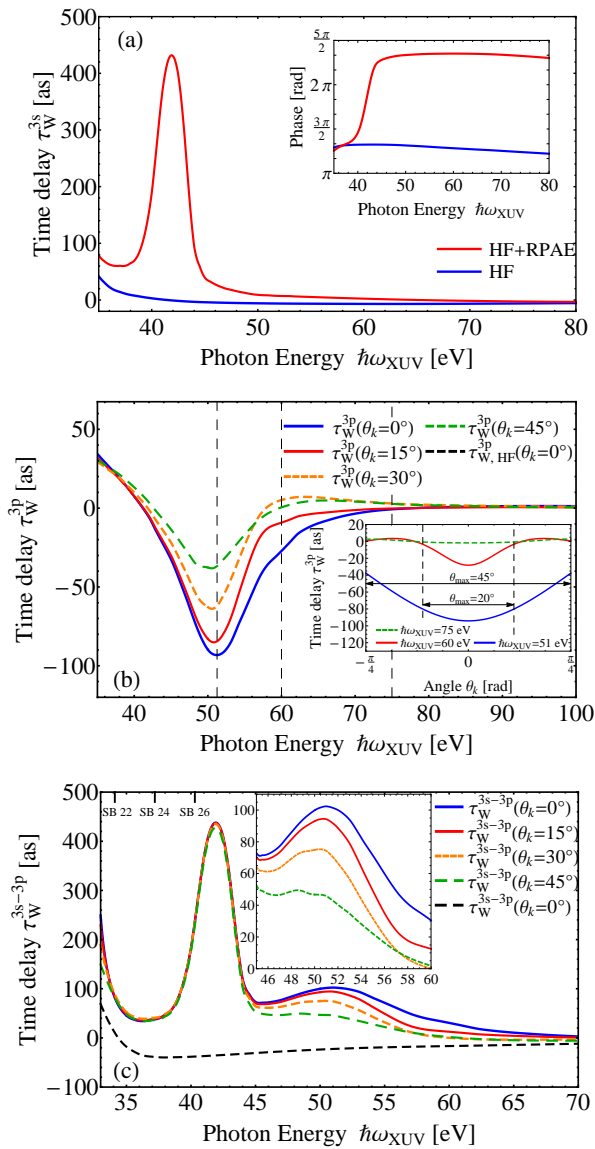}
\caption{(Color online) (a) Time delay corresponding to $3s$ photoionization. The inset show the scattering phases of the HF and RPAE methods. (b) Time delays $\tau_{\rm W}^{3p}$ in dependence on the photon energy $\hbar\omega_{\rm XUV}$ for different asymptotic directions $\theta_{\boldsymbol{k}}$ of the photoelectron. The inset shows the angular dependence for three different photon energies around the Cooper minimum. (c) Relative time delay $\tau_{\rm W}^{3s-3p}$ for different photon energies is shown. SB 22 - SB 26 mark the photon energies of the experiments \cite{Guenot2012,Klunder2001}.}
\label{DelayArgon}
\end{figure}

In case of argon we used the same shape and field amplitude of the electromagnetic perturbations. To reach improved results we used experimental bound state energies instead of the HF eigenvalues \cite{Kheifets2013}. Thus the energy difference between the $3s$ and $3p$ states is 13.48~eV \cite{Nist2}. To avoid accidental photoionization of both initial states we had to use a pulse with a longer duration in comparison to neon, which means that the energy spectrum is narrower. In case of argon a FHWM of 300~as is sufficient.\\
The panel~(a) of fig.~(\ref{DelayArgon}) shows the angular independent time delay corresponding to photoionization from the $3s$ initial state.  Electronic correlations as implemented in RPAE have marked effects that show up in the time delay. In particular the presence of the well-known Cooper minimum (in contrast to Ne) results in a distinctive feature \cite{Kheifets2013}. This Cooper minimum can not be reproduced by a SAE calculation. Thus the full numerical simulation with a SAE model potential to achieve the time delay information of argon is not performed.\\
In panel~(b) of fig.~(\ref{DelayArgon}) the time delay of the $3p$ photoionization is depicted, which is calculated according to eq.~\eqref{AveragedWD}. The expected Cooper minimum around the photon energy of 50~eV \cite{Cooper,PhysRevA.83.053401} is correctly reproduced.
Substantial variations in the time delay with the photon energy as well as a distinctive angular dependence are observed. We find a less pronounced peak for the $3p$ subshell in comparison to the $3s$ subshell (transitions to the final states with $\ell=0$ and $\ell=2$ are possible in the former case). This can be explained by the interference between the normally weak transition $\ell_{\rm i}=1\rightarrow\ell=0$, which become stronger near the Cooper minimum, and the normally dominant $\ell_{\rm i}=1\rightarrow\ell=2$ transition \cite{ArtFano}. Thus, the resulting time delay does not fall below -100~as. The discrepancies between the various energy-dependent curves for different asymptotic directions $\theta_{\boldsymbol{k}}$ vanish for larger photon energies.\\
In the region around the Cooper minimum the time delays corresponding to photoionization from $3p$ subshell are negative despite the positive energy derivative of the HF scattering phase of the dominant $\ell_{\rm i}=1\rightarrow\ell=2$ transition. This is due to the additional phase through RPAE correction that contributes with a negative $\pi$-jump at the Cooper minimum producing a local and very distinctive negative time delay \cite{Kheifets2013}. For larger angles, the peaks around the Cooper minimum become less distinctive. This characteristics can be explained by decreasing values of the spherical harmonics $Y_{20}$ (cf. with eq. \eqref{eq:DME_phase}) with increasing $\theta_{\boldsymbol{k}}$, which weakens transition $\ell_{\rm i}=1\rightarrow\ell=2$ further and increases the influence of the scattering phases $\delta^{\rm HF}_{\ell=0}+\delta^{\rm RPAE}_{\ell=0}$. The inset in fig.~\ref{DelayArgon}(b) shows the time delay of the $3p$ photoionization process angle-resolved for specific photon energies. In case of neon the time delay was nearly independent on the angle in the range between $-45^\circ$ and $45^\circ$ (cf. fig.~\ref{DelayNeon}(b)), while in the case of argon we find a strong angular variation at energies in the vicinity of the Cooper minimum (50~eV and 60~eV). For higher energies the situation is similar to the case of neon.\\
In fig.~\ref{DelayArgon}(c) the full relative delay $\tau^{3s-3p}_{\rm W}$ in dependence on the photon energy for different asymptotic directions is depicted. The graph reveals a large peak around 42~eV, which originates from the behavior of the $3s$ delay at the Cooper minimum. Characteristic for this feature is a very weak angular dependence. The larger variations with angle, originating from the $3p$ contribution, show up primary at larger photon energies.
The results for the forward direction, i.e. $\theta_{\boldsymbol{k}}=0^\circ$, are in good agreement with the calculations of Dahlstr\"om \emph{et al.} \cite{Dahlstrom2} and Kheifets \cite{Kheifets2013}.
We marked the photon energies corresponding to the findings of the interferometric experiments \cite{Klunder2001,Guenot2012} and see that for these energies there is no significant angular dependence of the relative delay, which could influence the result of the measurement.\\
In contrast at the photon energy of 51~eV the value of the relative time delay in forward direction amounts to 105~as, while an acceptance angle of $20^\circ$ results in an averaged relative delay of 92~as. A larger acceptance angle of $45^\circ$ leads to a relative time delay, averaged over all possible photoemission directions, of 68~as.

\section{Conclusion}

In the present work, we studied the angular dependence of the photoemission time delay from the valence shells of neon and argon. Results were obtained within the Random Phase Approximation with Exchange to include intershell correlations. We found that the existence of a Cooper minimum has a strong impact on the angular dependence of time delay. From an experimental point of view the angular dependence has almost no relevance in the case of neon because the effect is nearly not existent around the forward direction. In the case of argon for photon energies around the Cooper minimum of the $3p$ photoionization the angular dependence is distinctive and can have a sizable influence on the measured delay in this region.

\section*{Acknowledgement}
The authors thank Eleftherios Goulielmakis and Vladislav Yakovlev for enlightening discussions regarding the experimental setup and theory details.

\bibliographystyle{apsrev}
%\bibliography{v1}

\end{document}